\documentstyle[12pt]{article}
\def\lsim{\mathrel{\raise.2ex\hbox{$<$}\hskip-.8em\lower.9ex\hbox{$\sim$}}}
\def\gsim{\mathrel{\raise.2ex\hbox{$>$}\hskip-.8em\lower.9ex\hbox{$\sim$}}}
\def\mev{{\rm MeV}}

\def\km{{\rm km}}
\def\erg{{\rm erg}}
\def\mpc{{\rm Mpc}}
\def\m{{\rm m}}
\def\msec{{\rm msec}}

\def\j{{\rm J}}
\def\cm{{\rm cm}}
\def\khz{{\rm kHz}}

\begin{document}

\font\fortssbx=cmssbx10 scaled \magstep2
\hbox to \hsize{
\hbox{\fortssbx University of Wisconsin - Madison}
\hfill$\vcenter{\hbox{\bf MADPH-96-929}
                \hbox{January 1996}}$ }

\vspace{.5in}

\begin{center}
{\Large\bf Neutrinos from Gamma Ray Bursts}\\[4mm]
F. Halzen and G.  Jaczko\\
\it Physics Department, University of Wisconsin, Madison, WI 53706 USA
\end{center}

\let\Large=\large
\vspace{.75in}

\begin{abstract}
We show that the detection of neutrinos from a typical gamma ray burst requires
a kilometer-scale detector. We argue that large bursts should be visible with
the neutrino telescopes under construction. We emphasize the 3 techniques by
which neutrino telescopes can perform this search: by triggering on i) bursts
of muons from muon neutrinos, ii) muons from air cascades initiated by high
energy gamma rays and\break
iii) showers made by relatively low energy ($\simeq 100\,\mev$) electron
neutrinos. Timing of neutrino-photon coincidences may yield a measurement of
the neutrino mass to order $10^{-5}$~eV, an interesting range in light of the
solar neutrino anomaly.
\end{abstract}

\thispagestyle{empty}
\newpage

\section{Introduction}

The origin of gamma ray bursts (GRBs) is arguably astronomy's most outstanding
puzzle\cite{fishman}. Contributing to its mystery is the failure to observe
counterparts in any other wavelength of light. It should therefore be a high
priority to establish whether GRBs emit most of their energy in
neutrinos\cite{pacz,plaga,learned} as expected in the (presently favored)
cosmological models.

It is not the purpose of this paper to study the modelling of GRBs. We will
consider two cosmological scenarios: ultra-relativistic
fireballs\cite{piran} and cosmic strings\cite{babul} and reduce their
predictions to dimensional analysis, omitting
details which represent at best unfounded speculations. After imposing
experimental constraints on the dimensional analysis, it suffices to
quantitatively frame the question of neutrino emission. The ``experimental
facts", which will later constrain our model parameters, can be encapsulated as
follows\cite{plaga}: i) there are about 100 bursts per year with an
average fluency in photons of $F_\gamma \gsim10^{-9}\rm\,J\,m^{-2}$, ii) they
are concentrated, on average, at a redshift of $z\,\simeq\,1$, iii) some
bursts last less than 10~s, and iv) they do not repeat on a time-scale of
1~year or less.  Our predictions will be presented in a form
in which they can be scaled to fit varying interpretations of the experimental
situation. Our interpretation of the observational situation, as well as the
models presented, seem to be currently favored, although there are some
dissenters. For example, some advocate that the origin of GRBs can be traced to
an extended halo population of neutron
stars. However, the predictions of such models for neutrino emission may in the
end differ only slightly, since the reduced luminosity, compared to
large-redshift
sources, is compensated for by a reduced distance to the source.

\looseness=-1
Our results can be summarized as follows. The detection of typical GRBs
requires kilometer-scale neutrino telescopes. GRBs provide us with yet another
example of Nature's conspiracy to require kilometer-size detectors for
exploring our science goals\cite{fh-snowmass}, from dark matter searches to the
study of active galaxies. Rare, large bursts may however be within reach of the
present experiments. Our results will demonstrate that non-observation will
lead to meaningful constraints on the models. In particular, it is unlikely
that cosmic string models can escape the scrutiny of the detectors presently
under construction, because they predict a fluency in neutrinos which exceeds
that for photons by a factor of order $10^8$ or more.

Furthermore, we will emphasize the 3 techniques by which neutrino telescopes
can
search for GRBs. All detectors\cite{PR}, such as the DUMAND and NESTOR deep
ocean experiments, can search for short bursts of high energy muons of
$\nu_{\mu}$-origin. Sensitivity is good, i.e. atmospheric backgrounds small,
because the signal integrates over very short times and does not have to be
searched for; one looks at times given by the gamma ray observations. The
shallower detectors like AMANDA and Baikal can also search for the muons made
in air showers initiated by TeV gamma rays\cite{HS} of GRB origin. Finally,
AMANDA can use its supernova trigger\cite{hjz} to identify excess counting
rates in the optical modules associated with a flux of MeV-GeV $\nu_e$'s for
the duration of a gamma ray burst.

It has not escaped our attention that the observation of coincident bursts of
neutrinos and gamma rays can be used to make a measurement of the neutrino
mass. The mass is determined from the time delay $t_d$ by simple relativistic
kinematics with $m_{\nu}=E_{\nu}\sqrt{2c\,t_d/ D}$. With $t_d$ possibly of
order milliseconds, distances $D$ of thousands of Megaparsecs and energies
$E_{\nu}$ similar to that of a supernova, neutrino observations from GRBs could
improve the well-advertised limit obtained from supernova SN1987A by a factor
$10^6$. The sensitivity of order $10^{-5}$~eV is in the range implied by the
solar neutrino anomaly. The measurement would be greatly facilitated by the
fact that, unlike for rare supernova events, repeated observations are
possible.

\section{Accelerator I: The Relativistic Fireball Scenarios}

Although the details can be complex, the overall idea of fireball models is
that a large amount of energy is released in a compact region of radius
$R\simeq 10^2$\,km$\simeq c \Delta t$. The shortest time-scales, with $\Delta
t$ of order milliseconds, determine the size of the initial
fireball\cite{piran}. Only neutrinos escape because the fireball is opaque to
photons. In GRBs a significant fraction of the photons is indeed above pair
production threshold and produce electrons. It is straightforward to show that
the optical depth of the fireball is of order $10^{13}$\cite{piran}. It is then
theorized
that a relativistic shock, with $\gamma \simeq 10^2$ or more, expands into the
interstellar medium and photons escape only when the optical depth of the shock
has been sufficiently reduced. The properties of the relativistic shock are a
matter of speculation. They fortunately do not affect the predictions for
neutrino emission.

For a fluency $F=10^{-9}\rm\,J\,m^{-2}$ and a distance $z=1$ the energy
required is
\begin{equation}
E_\gamma = 2\times 10^{51}{\rm\, erg} \left(D\over 4000{\rm\; Mpc}\right)^2
\left(F\over 10^{-9}\rm\,J\>m^{-2}\right) \,,  \label{eq:E_gamma}
\end{equation}
using $E_\gamma = 4\pi D^2 F$. The temperature $T_\gamma$ is obtained from the
energy density
\begin{equation}
\rho = {E_\gamma\over V} = {1\over2} h a T^4 \,,
\end{equation}
where $h$ represents the degrees of freedom ($h_\gamma = 2$ and $h_\nu =
2\cdot3\cdot{7\over8}$ for 3 species of neutrinos and antineutrinos), $V$ the
volume corresponding to radius $R$ and $a = 7.6 \times 10^{-16}\rm\; J\; m^{-3}
\,K^{-4}$. We find that
\begin{equation}
T_\gamma = 8\;\mev \left(E_{\gamma}\over 2\times10^{51}\,\erg\right)^{1/4}
\left(100\;\km \over R\right)^{3/4} \,. \label{eq:T_gamma}
\end{equation}
For neutrinos
\begin{equation}
T_\nu = \left(E_\nu/E_\gamma\over h_\nu/h_\gamma\right)^{1/4} T_\gamma \,.
\label{eq:T_nu}
\end{equation}
For a merger of $n$-stars, for instance, the release of a solar mass of energy
of $2 \times 10^{53}$~erg implies a total energy emitted in neutrinos
$\sim10^2E_\gamma$. The $\gamma$'s are most likely produced by bremsstrahlung
of electrons from $\nu\bar\nu$ annihilation. The actual predictions for the
energy and time structure of the photon signal depend on the details of the
shock which carries them outside the opaque fireball region of size $R$. The
data suggest that the structure of these shocks is complex. Neutrinos, on the
contrary, promptly escape and carry direct information on the original
explosion. From (\ref{eq:T_gamma}),(\ref{eq:T_nu}) we obtain $T_\nu\simeq2.5
T_\gamma\simeq20\;\mev$. Using this and a total neutrino energy in the fireball
of $10^2E_\gamma$ we obtain
\begin{eqnarray}
E_{\nu} &=& 3.15\ T_{\nu} \ =\ 65\;\mev \left(E_\gamma\over
2\times10^{51}\,\erg\right)^{1/4} \left(100\;\km\over R\right)^{3/4} \,,
\label{eq:E_nu_obs}\\
\Delta t_{\rm obs} &=& 0.3\;\msec \left(R\over 100\;\km\right) \,.
\end{eqnarray}
The neutrino fluency is obtained from $E_{\nu\rm\,tot} / (4\pi D^2)$
\begin{equation}
N_\nu = 10^4\,\m^{-2} \left(E_{\nu\rm\,tot}\over2\times10^{53}\,\erg\right)
\left(65\;\mev\over E_\nu\right) \left(4000\;\mpc\over D\right)^2
\end{equation}
or more than $10^{57}\ \nu$'s at the source. Notice that this prediction is
rather model-independent because it just relies on the fact that a solar mass
of energy is released in a volume of 100~kilometer radius which is determined
by the observed duration of the bursts.

Although the $\sim100\,\mev$-neutrinos are below the muon threshold of high
energy
neutrino telescopes, the $\bar\nu_e$ will initiate electromagnetic
showers by the reaction  ($\bar\nu_e + p \to n + e^+)$ which will be counted by
the AMANDA supernova trigger.

A supernova with properties similar to those of SN1987A can cause a 10 second
burst
of neutrinos in the AMANDA detector with $E_{\nu} \simeq 40$~MeV. They produce
positrons with, on average, half that energy. Detailed
simulations\cite{hjz} of the supernova signal in the AMANDA detector have shown
that each photomultiplier tube (PMT) has a seeing radius $d\simeq7.5$~m for
20~MeV positrons. The number of events per PMT is given by
\begin{equation}
\#{\rm N_{\nu\,\rm obs}} \simeq N_{\nu} (\pi d^2) \left(d\over \lambda_{\rm
int} \right) \,. \label{eq:events}
\end{equation}
The last factor estimates the probability that the $\bar\nu_e$ produces a
positron within view of the PMT. Here
\begin{equation}
\lambda_{\rm int}^{-1} = {2\over18} A \rho \sigma_0 E_{\nu}^2
\end{equation}
with
\begin{equation}
\sigma_0 = 7.5\times10^{-40}\rm\,m^2\,MeV^{-2} \,.
\end{equation}
$A$ is Avogadro's number and $\rho$ the density of the detector medium. One
should not forget here that the dependence of the cross section on neutrino
energy is linear rather than quadratic above $\sim100\,\mev$.

We have checked by Monte Carlo\cite{zas} that the seeing volume scales linearly
in the energy of the positron, or neutrino, up to TeV energies. Eventually the
radius will cease to grow due to attenuation of the light. With absorption
lengths of several hundred meters\cite{Rome} this upper
limit is outside the range of where we will apply (\ref{eq:events}).
Therefore, the event
rate for GRBs is given by (\ref{eq:events}) with $d =
7.5\;\m\left(30\;\mev\over20\;\mev\right)^{1/3}$. Here $30\,\mev$ is the
positron energy which is, on average, half the neutrino energy given by
Eq.~(\ref{eq:E_nu_obs}).

Can this signal be detected by simple PMT counting? Signal $S$, noise $N$ and
$S/\sqrt N$, for an average burst, are given by
\begin{eqnarray}
S &=& 10^{-3}{\rm\;events} \left(N_{\nu\rm\, obs}\over 5 \times 10^{-6}\right)
\left(D_{\rm PMT} \over 20\;\cm\right)^2 \left(N_{\rm PMT}\over 200\right)
\,,\\
N &=& 60{\rm\;events} \left(\Delta t\over 0.3\;\msec \right) \left(N_{\rm
back} \over 1 \khz\right) \left(N_{\rm PMT}\over 200\right) \,,\\
S/\sqrt N &=& 10^{-4} \left(N_{\rm back}\over 1\;\khz\right)^{-1/2}
\left(D_{\rm PMT}\over 20\;\cm\right)^2 \left(N_{\rm PMT}\over200\right)^{1/2}
\,.
\end{eqnarray}
AMANDA has been chosen for reference with 200 PMTs with a diameter $D_{\rm
PMT}$ of 20~cm and a background counting rate of roughly 1~kHz. With such low
rates in millisecond times, observation obviously requires a dedicated trigger.

Obviously the event rate for an {\it average} burst is predicted to be low. We
will argue nevertheless that observation is possible and clearly guaranteed
for kilometer-scale detector with several thousand PMTs. First, the parameters
entering the calculation are uncertain. The event rate increases with neutrino
energy as $E_{\nu}^3$ because of the increase of the PMT seeing distance $d$
and the neutrino interaction cross section $\sigma_0$. With increased energy
the average burst may become observable. Individual burst can yield orders of
magnitude higher neutrino rates because of intrinsically
higher luminosity and/or smaller than average distance to earth. For example, a
burst
10 times closer than average and 10 times more energetic is observable with a
significance of well over 10~$\sigma$ in the exisiting AMANDA detector. Given
the uncertainties in the model and its parameters as well as the chaotic nature
of the phenomenon (there is no such thing as an average GRB), this event
represents a plausible possibility.

As demonstrated by the $\gamma$-ray observations, the structure of the shock
producing the gamma rays is complex. The interaction of multiple shocks can
also produce neutrinos on other time-scales and with different, sometimes much
higher, energies\cite{pacz}. So one should have an open mind when searching for
bursts. This is underscored by the rather different predictions obtained from
string-type models, which we discuss next.

\section{Accelerator II: Cosmic String-Type Scenarios}

The dimensional analysis relevant to accelerators such as cosmic strings is
synchrotron emission from a beam of ultra-relativistic particles. The time of
emission is now given by
\begin{equation}
\Delta t_{\rm lab} = {L\over c\gamma^3} \,.
\end{equation}
Here $L$ is the size of the accelerator and $\gamma=I_{\rm saturation}/I$ is a
ratio of electric currents, which is some large number. One main difference
with the previous scenario is that the emission is relativistically beamed in a
solid angle of size $\gamma^{-2}$. The idea is that when accelerated currents
reach a value $I_{\rm saturation}$ it is energetically more favorable to
radiate away the mass of the accelerating cosmic source, rather than sustain
the high current. This happens for instance at cusps in oscillating loops where
the current becomes, theoretically, infinitely large. A mass $\mu$ per unit
length $L$ is radiated away in a time $\Delta t$. In dimensionless units, $\mu$
is,
\begin{equation}
\epsilon = \mu{G\over c^2} \,.
\end{equation}
A dimensional estimate for $L$, the size of the cosmological accelerator, can
be made as follows. The time over which a cosmic accelerator loses mass is
clearly proportional to $L/\mu$ or, in correct units, $L/\epsilon c$. We equate
this to the only time in the problem: the lifetime of the universe at the
redshift of the accelerator,
\begin{equation}
{L\over \epsilon c} = \xi\; {t_0\over (1+z)^{3/2}} \,, \label{eq:time}
\end{equation}
where $ct_0 = 6 \times 10^{27}\,\cm$ and the proportionality factor $\xi=1$. So
$L= \xi \epsilon ct_0/(1+z)^{3/2}$ and we can now calculate the duration of the
burst
\begin{equation}
\Delta t_{\rm observ} = (1+z) \Delta t_{\rm lab}
= (1+z) {L\over c\gamma^3} = 10^{17}\xi\;{\epsilon\over\gamma^3}
\;\rm seconds\,.
\end{equation}
In the accelerator frame (comoving frame)
\begin{equation}
\Delta t_{\rm com} \cong \xi\; {\left(\epsilon / 10^{-11} \right) \over
\left( \gamma / 10^3 \right)^2} \rm\; seconds \,,
\end{equation}
The choice of units will become clear further on. The energy loss per unit
length is independent of $\epsilon$ with
\begin{equation}
{\mu c^2\over \Delta t_{\rm com}} = {1 \over \xi}\; 8\times 10^{33}
\left(\gamma\over10^3\right)^2 \rm\, J\, m^{-1}\, s^{-1} \,. \label{eq:loss}
\end{equation}
A fraction $\eta_\gamma$ is radiated away in $\gamma$-rays.

The above equations are valid for cosmological strings or loops of false vacuum
in grand unified theories. Near cusps in oscillating loops the particle
currents become very large, creating a situation where the energy density
exceeds that of the topological defect and the energy is released in a short
localized burst of radiation. In string models there is a proportionality
factor multiplying the r.h.s.\ of (\ref{eq:time}) which is of order $\xi=10^3$
rather than unity; see e.g.\ Ref.~\cite{babul}. From now on we will include
this factor, so that our results can be directly compared to these models.

Imposing the ``experimental facts", listed in the introduction, on the
dimensional analysis (with $\xi=10^3$) yields the following
constraints\cite{plaga,babul}:
\begin{eqnarray}
&10^2  < \gamma < 10^5  &\nonumber\\
&10^{-12}  < \epsilon <  10^{-11}& \\
&10^{-10}  < \eta_\gamma  <  10^{-9} &\nonumber
\end{eqnarray}
The critical result here is that to accommodate the time-scales as well as the
fluencies in a large redshift source of this type, the fraction of energy loss
into gamma rays is actually very small, $10^{-10}$ to $10^{-9}$. Theoretical
arguments\cite{plaga} lead to the expectation that most of the energy is
radiated into $\nu$'s. This fits well with the observational fact that the
missing energy is not emitted in any other wavelength of light.

Before proceeding it is important to point out that the small fraction of the
burst energy going into gamma rays is not a surprise. Cosmic strings belong to
the class of highly inefficient models in which the whole accelerator is
boosted
by a Lorentz factor $\gamma$. In contrast, conventional fireball models
describe a collisionless shock of protons which carries kinetic energy far
outside the opaque fireball where
it is transformed into a burst of photons.

A fraction $\eta_\gamma^{-1}$ is radiated into $\nu$'s of energy $E_{\nu\,\rm
obs}$. The flux for a typical burst is
\begin{equation}
N_\nu = {1\over\eta_\gamma} \, {10^{-9}\,\rm J\, m^{-2} \over E_{\nu\rm\,obs}}
\end{equation}
or
\begin{equation}
N_\nu {\rm\ per\ cm^2} = 10^8 \left(\eta_\gamma \over 10^{-10}\right)^{-1}
\left(E_{\nu\rm\,obs} \over 100\rm\ MeV\right)^{-1} \left(F_\gamma\over
10^{-9}\rm\,
J\, m^{-2} \right)
\end{equation}
during a time
\begin{equation}
\Delta t_{\rm obs} = {1\over\gamma}\Delta t_{\rm com} = 1{\rm\;sec}
\left(\epsilon\over 10^{-11}\right) \left(\gamma\over10^3\right)^{-3} \,.
\end{equation}
Here
\begin{equation}
E_{\nu\rm\,obs} = \gamma \, 3.15\;T_{\nu\rm\,com} \,.
\end{equation}
The thermal emission of the neutrinos in the accelerator frame follows a
Fermi-Dirac distribution with temperature $T_{\nu\rm\,com}$. We will estimate
it next following Ref.~\cite{plaga}.

Consider an accelerator segment of loop of length $L$ and radius $R$. Assume
black body radiation off its surface and apply the Stefan-Boltzmann law in a
comoving  frame. Using~(\ref{eq:loss}),
\begin{equation}
{\mu c^2\over \Delta t_{\rm com}} \, L = (2\pi RL) (\sigma T_{\nu\,\rm com}^4)
\,,
\end{equation}
where $\sigma$ is the Stefan-Boltzmann constant. We obtain
\begin{equation}
T_{\nu\,\rm com} = {E_{\nu\rm\,obs}\over3.15\;\gamma} = (10{\rm\ MeV})
\left(\gamma\over10^3\right)^{1/2} \left(10^{-7}\,{\rm m}\over R\right)^{1/4}
\,.
\end{equation}
For a cosmic string $R=I_{\rm saturation}/H_{\rm cr}$, where $H_{\rm cr}$ the
critical field
strength. $I_{\rm saturation}$ was calculated by Witten\cite{witten}, and is
typically
\begin{equation}
10^{-8} {\rm m} < R < 10^{-6} {\rm m} \,.
\end{equation}

The possibilities covered by this class of  models range from thermal
supernova-type energies to TeV-neutrinos.  For illustration, we show results
for a low and high energy neutrino scenario.
\[
\begin{array}{c}
\gamma = 10^2 \qquad R = 10^{-6} \qquad E_{\nu\,\rm obs} = 560\rm\ MeV\\
2\times 10^7 < N_{\nu\,\rm obs} < 2\times10^8\rm\, per\ cm^2\\
10^2 < \Delta t_{\rm obs} < 10^3\rm\, seconds
\end{array}
\]
or
\[
\begin{array}{c}
\gamma = 10^5 \qquad R = 10^{-8} \qquad E_{\nu\,\rm obs} = 60\rm\ TeV\\
2\times 10^2 < N_{\nu\,\rm obs} < 2\times10^3\,\rm per\ cm^2\\
0.1 < \Delta t_{\rm obs} < 1\rm\ \mu sec
\end{array}
\]

Suppose neutrinos with $E_{\nu\,\rm obs} \simeq 40$~MeV produce electrons in
the detector with energy $(1-\langle y\rangle)E_\nu$, or about 20~MeV, just
like SN1987A would have produced in AMANDA. We calculate a flux of
$5\times10^8$
per cm$^2$ in a rather long burst. We know from the supernova analysis that
each PMT has a seeing radius $d\simeq7.5$~m in this case. The number of events,
given by (\ref{eq:events}), is 10 per PMT for a typical, {\rm average} burst.
This is 10 times smaller than a supernova, but the GRB data indicates that we
have 100 shots per year and
there should be some big ones. Models suggest searches over $>1$~sec intervals,
maybe up to 1000~sec. Also notice that event rates grow with energy as
$\sigma d^3/E$. Both $d,\sigma$ grow with energy. The signals should be
spectacular for $E_{\nu\rm\,obs}$ values of hundreds of MeV or more.

An extreme example on the high energy end yields $\sim10^2$ neutrinos of tens
of TeV
energy per cm$^2$ in periods $\ll 1$~sec. In this scenario, the secondary muons
can be detected
and reconstructed. This allows one to both count the neutrinos and reconstruct
their direction with degree-accuracy. The event rates are now given
by\cite{PR}:
\begin{eqnarray}
{\rm N_{events}} &=& N_\nu \,{\rm Area} \, P_{\nu\to\mu} \,,\\
P_{\nu\to\mu} &\simeq& \rho \sigma_\nu R_\mu = \rho \left( 10^{-42}\,{\rm m^2}
\, {E_\nu\over{\rm GeV}}\right) \, \left( 5{\rm\,m} \, {E_\mu\over{\rm GeV}}
\right)\,.
\end{eqnarray}
Here $P_{\nu\to\mu}$ is the probability that the neutrino interacts and spawns
a muon that reaches the detector; it is proportional to the density $\rho$ of
the detector medium, the neutrino interaction cross section $\sigma_\nu$ and
the muon range $R_\mu$. For $E_\mu\simeq{1\over2}E_\nu\simeq30$~TeV and
$\rho={11\over18}A$ per cm$^3$ we have $P_{\nu\to\mu}= 10^{-3}$ or 10$^5$
events for a detector as small as 100~m$^2$ area detector!

Therefore, bursts associated with topological defects are unlikely to escape
the scrutiny
of  both the supernova and the muon trigger. In part of the parameter space one
should be able to rule out the
cosmological models even for average bursts. In other regions, one can
constrain the models only from a search for energetic bursts.

\section{Detecting $\gamma$-Rays with Neutrino Telescopes?}

What about seeing $\gamma$-rays? Shallow detectors like AMANDA and Baikal
detect secondary muons produced by $\gamma$-showers in the atmosphere. For a
vertical muon threshold of 180~GeV, AMANDA should be sensitive to TeV gamma
rays. The number of photons is calculated from the fluency $F_\gamma$ by
\begin{equation}
N_\gamma(>E) = {1\over\alpha} {F_\gamma\over E_\gamma^\alpha} \,,
\end{equation}
where $\alpha$ is the spectral index ($\alpha=1$ for Fermi shocks). For
$\alpha=1$ and a fluency per burst of $10^{-9}\,\j\;\m^{-2}$ we find that
$F_\gamma =10^{-2}  \ln^{-1}\!\left(E_{\gamma\,{\rm max}} \over E_{\gamma\,{\rm
min}} \right)$ per m$^2$ per burst. There is a rather weak logarithmic
dependence on the maximum and minimum energy of the photons in the burst.
Notice that the TeV flux, even if it exists, is too small to be detected by
satellite experiments. The maximum energy of GRBs is therefore an open
question. It has been speculated that they may be the sources of the highest
energy cosmic rays which implies a very high energy accelerator indeed.

The muon flux produced by above gamma ray flux can be computed following Halzen
and Stanev\cite{HS}:
\begin{eqnarray}
N_\mu ({>}E_\mu) &\simeq& 2\times 10^{-5} {F_\gamma\over \cos\theta}
{1\over (E_\mu/\cos\theta)^{\alpha+1}}\, \ln\!\left(\cos\theta E_{\gamma\,{\rm
max}} \over 10E_\mu\right) \left(E_\mu/\cos\theta\over 0.04\right)^{0.53}\;.
\end{eqnarray}
Here $E_{\mu}$ is the vertical threshold energy of the detector, e.g.\ 0.18~TeV
for the AMANDA detector. $\theta$ is the zenith angle at which the source is
observed. This parametrization reproduces the explicit Monte Carlo results.

We predict 10$^{-6}$ muons per m$^2$ for an average burst, which can therefore
be detected in a km$^2$ telescope! The probability that a 1~TeV $\gamma$
contains a detectable muon is about 10$^{-4}$. We assumed here a burst in the
1~MeV to 10~TeV range and $\cos\theta = 1$. All this requires, of course, that
the GRB flux extends to TeV energies. We do not know whether any do because
satellite experiments have no sensitivity in this energy range. There is no
atmospheric $\mu$ background in a pixel in the sky containing the GRB on a 1
second time scale. Big bursts
may be detectable in the $10^4\,\m^2$ detectors presently under construction.

\section*{Acknowlegdments}
We thank M. Drees, J. Jacobsen, M. Olsson, R. Plaga and E. Zas for discussions.
This research was supported in part by the U.S.~Department of Energy under
Grant No.~DE-FG02-95ER40896 and in part by the University of Wisconsin Research
Committee with funds granted by the Wisconsin Alumni Research Foundation.

\end{document}